\documentclass[runningheads,a4paper]{llncs}
\usepackage[T1]{fontenc}
\usepackage{graphicx}
\usepackage[numbers]{natbib}
\usepackage{multirow}
\usepackage{booktabs}
\graphicspath{{figures/}}
\usepackage{amsmath,amssymb}  
\usepackage{tabularx}  
\usepackage{enumitem}
\usepackage{color}
\usepackage{hyperref}         

\usepackage{tikz}
\usetikzlibrary{arrows.meta, positioning}
\definecolor{regimeGold}{HTML}{B8941F}
\definecolor{regimeSilver}{HTML}{6E7782}
\definecolor{regimeMix}{HTML}{3D6FA5}
\definecolor{regimeCohort}{HTML}{8A8A8A}
\tikzset{
  regnode/.style={
    rectangle, rounded corners=4pt, draw=#1, line width=0.9pt, fill=#1!12,
    align=center, inner sep=6pt, minimum height=10mm, font=\footnotesize,
  },
  regedge/.style={
    draw=#1, line width=0.9pt, -{Stealth[length=4.5pt, width=4pt]},
    shorten >=1.5pt, shorten <=1.5pt,
  },
}
\begin{document}
\title{False Confidence: Automated Labels Confound Fairness Audits in Cervical Spine Segmentation}
\titlerunning{Fairness in Cervical Spine MRI Segmentation}
%
\author{Linus Juni \and Aasa Feragen \and Aditya Parikh}
\authorrunning{L. Juni et al.}
\institute{Section for Visual Computing, DTU Compute,\\
Technical University of Denmark, Kongens Lyngby, Denmark\\
\email{\{s225224, afhar, adipa\}@dtu.dk}}

\maketitle              
\begin{abstract}

Automated segmentation of cervical-spine MRI is increasingly used in clinical workflows, yet no fairness audit exists for this
anatomy. We show that auditing these segmentation tasks is complicated by a common property of modern segmentation datasets: expert-annotated gold labels are expensive, so abundant machine-generated (silver) labels are added to limit annotation cost. This matters because the reference used to judge a model can itself be biased. In this study, we present the first fairness audit of cervical-spine MRI segmentation across sex, age, and race using the CSpineSeg dataset. We observe that the deployed model is demographically fair, but the choice of reference label, however, is not neutral. Because a dataset's silver labels are generated by a model
trained on its gold labels, any new model trained on those same gold labels agrees more
with the silver labels than with expert truth: scoring identical predictions against
silver rather than gold overestimates performance by ${\sim}8$ Dice points and turns
the fairness verdict for age from non-significant to significant -- not by the gap
inflation Parikh et al.\ report (which we term \emph{false magnitude}) but by collapsing
within-group variance (which we term \emph{false confidence}). Reference-label provenance
is thus a first-order confounder in segmentation evaluation: performance and fairness
should be reported against expert labels, and any fairness claim stated together with
the provenance of its reference.

\end{abstract}

\section{Introduction}

Automated segmentation of the cervical spine -- labelling the vertebral bodies and
intervertebral discs on MRI helps clinicians grade stenosis and track degeneration
over time~\cite{kato2012cervical}. A model that works well for some patient groups and produces incorrect segmentation for others would quietly widen existing gaps in patient care and surgical planning. Bias audits are increasingly required before clinical use, and are not optional~\cite{euaiact2024}. Yet no fairness audit exists for this anatomy.

Auditing a segmentation model is harder than auditing a classification model. Its output is millions of voxels rather than a single decision, so fairness is condensed to a single scalar (Dice, HD95) and the metric is itself biased by structure size when anatomy differs between groups~\cite{taha2015metrics,parikh2026fairnesslabelbiasimage}. Expert annotation is also time- and cost-intensive, so modern segmentation datasets increasingly mix a few expert (gold) labels with many machine-generated (silver) labels~\cite{tajbakhsh2020embracing}. This carries its own problems, and a biased ruler can mislead an audit in two opposite ways. Parikh et al.~\cite{parikh2025labelbias} showed the first in breast MRI (a
\textit{biased ruler effect}): a silver reference biased against a group
\emph{exaggerates} the true gap, so a disparity looks larger than it is (\emph{false
magnitude}). Carrying their framework to a new anatomy, we find the opposite mode: a
silver reference that is a near-copy of the model under audit agrees with it so tightly
that score variation collapses, making a tiny, negligible gap look statistically certain
(\emph{false confidence}).

Fairness in segmentation is much less studied~\cite{puyolanton2021fairness,danaee2026brain}. The only prior spine fairness study, FairMedFM~\cite{jin2024fairmedfm}, covers the lumbar spine, sex only, and uses no volume correction; FairSeg~\cite{tian2024fairseg} is in ophthalmology. None audit cervical-spine MRI across several protected attributes, and none ask whether the labels come from training or evaluation, and can, by themselves, carry bias.

\smallskip\noindent\textbf{Our Contributions.}\enspace Building on the biased-ruler
framework of Parikh et al.~\cite{parikh2025labelbias,parikh2025whofails}, we contribute:
\begin{enumerate}[label=\roman*]
  \item \textbf{The first demographic fairness audit of cervical-spine MRI
    segmentation}, across sex, race, and age -- on which the deployed model is fair
    by every standard metric.
  \item \textbf{Evidence that machine-generated labels leak from the expert labels
    they derive from.} Scoring identical predictions against silver rather than
    expert labels overestimates performance by ${\sim}8$ Dice points, affecting
    \emph{any} user who validates on such a dataset, not only fairness auditors.
  \item \textbf{A complementary mode of the biased-ruler effect.} From that same
    leakage, our correlated ruler flips age from non-significant to significant by
    collapsing within-group variance (\emph{false confidence}) -- the opposite
    mechanism to Parikh et al.'s \emph{false magnitude}~\cite{parikh2025labelbias}.
  \item \textbf{Practitioner guidance.} Report performance and fairness against
    expert labels, and treat a reference that scores both much higher and much more
    tightly than experts (${\sim}4\times$ variance collapse) as too close to the
    model to be an independent benchmark.
\end{enumerate}

\section{Dataset and Exploratory Analysis}
\label{sec:dataset}

\subsection{The CSpineSeg Cohort}
\label{sec:cohort}

We work with CSpineSeg~\cite{zhou2025cspineseg}, a public collection of 1,255
sagittal T2-weighted cervical-spine MRI exams from 1,232 patients at Duke
University. We drop a single localizer series with anomalous dimensions, leaving a
working set $\mathcal{D}_0$ of $N_0 = 1{,}254$ exams from 1,231 patients. The task
is semantic segmentation: labelling every voxel of a scan as background, vertebral
body, or intervertebral disc ($\mathcal{L} = \{0, 1, 2\}$).
About $40\%$ of the exams carry expert (gold) reference segmentations, drafted by one post-doctoral researcher without medical training and reviewed and revised by
one of six board-certified radiologists; the rest were segmented automatically by the authors' own model, and
these silver masks were never human-reviewed -- though the mid-sagittal slice of
each unannotated scan was checked for image quality.

We write the cohort as $\mathcal{D} = \{(X_i, Y_i, A_i, q_i)\}_{i=1}^{N}$: exam
$i$ pairs a 3D image $X_i$ with a reference segmentation $Y_i$, an attribute value
$A_i$ over the protected attributes $\mathcal{A} = \{\text{sex}, \text{race},
\text{age}\}$, and a provenance flag $q_i \in \{\mathsf{g}, \mathsf{s}\}$ marking a
gold (expert) or silver (machine-generated) label. We write
$\mathcal{D}_{A=a}$ for the subgroup with $A = a$. The analysis cohort
$\mathcal{D}$ ($N = 1{,}142$) is the sex-balanced subset of $\mathcal{D}_0$ defined
in ``\nameref{sec:regimes}''.

\autoref{tab:cohort} summarises $\mathcal{D}_0$: it leans slightly female, is
predominantly White with a sizeable Black minority, overwhelmingly non-Hispanic,
spans the adult age range, and splits across two manufacturers and two field
strengths. The volumes are strongly anisotropic -- high in-plane resolution
(${\sim}0.53$\,mm) but thick, few sagittal slices (${\sim}4$\,mm, 12--25 per scan).
These group sizes decide which axes we test formally -- sex (F/M), race
(White/Black), and age have enough exams per group for reliable inference --
while the smaller race categories (8\% combined) and ethnicity are too sparse and
reported descriptively only.

\begin{table}[t]
\centering
\caption{Composition of the working set $\mathcal{D}_0$ ($N_0 = 1{,}254$ exams,
counted at the exam level). Percentages are of $N_0$; age is mean\,$\pm$\,SD and the
13 missing-age exams (confirmed ${>}89$) are grouped into $\ge$60. The cohort is
overwhelmingly non-Hispanic ($92.7\%$); the $7.7\%$ Other/unknown race is mostly Asian
and unreported.}
\label{tab:cohort}
\small
\begin{minipage}[t]{0.48\textwidth}
\centering
\begin{tabular}{l l r}
\toprule
Characteristic & & $n$ (\%) \\
\midrule
\multirow{2}{*}{Sex}
  & Female & 683 (54.5) \\
  & Male   & 571 (45.5) \\
\addlinespace
\multirow{3}{*}{Race}
  & White         & 809 (64.5) \\
  & Black         & 349 (27.8) \\
  & Other/unknown & 96 (7.7)   \\
\addlinespace
\multirow{2}{*}{Field strength}
  & 1.5\,T & 746 (59.5) \\
  & 3.0\,T & 508 (40.5) \\
\bottomrule
\end{tabular}
\end{minipage}%
\hfill
\begin{minipage}[t]{0.48\textwidth}
\centering
\begin{tabular}{l l r}
\toprule
Characteristic & & $n$ (\%) \\
\midrule
\multicolumn{2}{l}{Age, years (mean $\pm$ SD)} & $54.6 \pm 16.3$ \\
\addlinespace
\multirow{3}{*}{Age group}
  & $<$40   & 241 (19.2) \\
  & 40--60  & 486 (38.8) \\
  & $\ge$60 & 527 (42.0) \\
\addlinespace
\multirow{2}{*}{Manufacturer}
  & Siemens & 788 (62.8) \\
  & GE      & 466 (37.2) \\
\bottomrule
\end{tabular}
\end{minipage}
\end{table}

\subsection{Anatomical Confounders}
\label{sec:confounders}

A performance gap is bias only if anatomy does not explain it, and structure size is the
main confound: men's vertebral bodies and discs are ${\sim}23$--$24\%$ larger, and the
vertebral body also grows with age as cervical degeneration
accrues~\cite{kato2012cervical} and differs by race, while the disc stays essentially
constant -- the least size-confounded cross-check. We therefore report vertebral body and
disc separately alongside their macro average, and score with a volume-aware nDSC
(``\nameref{sec:metrics}'') beside raw Dice so a size effect is not read as bias. The
size gap is anatomy, not scanner (GE vs.\ Siemens differ ${\sim}2.15\times$ in voxel count
yet ${\sim}6\%$ in mm$^3$), and no attribute is entangled with the scanner; only age is
mildly entangled with race (White patients slightly older), so we control for age when
comparing race.

\section{Methodology}
\label{sec:methodology}

Our design carries the label-bias and biased-ruler framework of Parikh et
al.~\cite{parikh2025labelbias,parikh2025whofails} from breast DCE-MRI over to
cervical-spine segmentation, with one structural difference that shapes everything
below: each CSpineSeg exam carries a gold or a silver label, never both.

\subsection{Data, Splits, and Label Regimes}
\label{sec:regimes}

The provenance flag $q_i$ partitions the cohort into disjoint gold and silver
sets, $\mathcal{D}^{\mathsf{g}}$ and $\mathcal{D}^{\mathsf{s}}$
($\mathcal{D}^{\mathsf{g}} \cap \mathcal{D}^{\mathsf{s}} = \varnothing$). We split
the cohort at the patient level into training, validation, and test sets in a
$70/10/20$ ratio, stratified by the protected attributes $\mathcal{A}$; splitting
by patient keeps
all of a multi-exam patient's exams on one side, preventing leakage. We then
balance the sexes -- the working set leans female (``\nameref{sec:cohort}''), and
an unequal base rate would conflate a true performance gap with mere
representation -- by randomly downsampling female exams within each split until the
two are equal, dropping $112$ exams ($N_0 = 1{,}254 \to N = 1{,}142$); this removes
representation as an explanation rather than attempting to close the
gap~\cite{parikh2025labelbias}. Because nnU-Net is cross-validation-native, we pool
$\mathcal{D}_{\mathrm{train}}$ and $\mathcal{D}_{\mathrm{val}}$ into a single
development set and hold out $\mathcal{D}_{\mathrm{test}}$.

We train two models that differ only in the labels they see, both sharing the
architecture, recipe, patient-level split, and stratification: the deployment-realistic
$M_{\mathrm{mix}}$ on the full split ($798$ exams: $318$ gold, $480$ silver) -- combining
scarce expert labels with abundant automated ones is the standard response to annotation
cost~\cite{tajbakhsh2020embracing} -- and a gold-only $M_{\mathrm{gold}}$ ($288$ exams),
which doubles as the generated-silver ruler for E2 (it never sees the test cases). Each
is sex-balanced independently within its split.

\subsection{Models, Metrics, and Experimental Design}
\label{sec:model}

\paragraph{Model and training.}
We segment every regime with nnU-Net v2 (ResEnc-L
preset)~\cite{isensee2021nnunet,isensee2024revisited}: a strong equal-resource baseline
in one fixed configuration shared across regimes, so every comparison is like-for-like
and any gap reflects the data and labels, not a weak model. Per regime we train the
standard 2D and 3D full-resolution configurations as five-fold cross-validations over
the pooled development set and ensemble the ten folds (twenty total); the one
audit-driven change is demographically stratified folds, with post-processing limited
to a global largest-foreground cleanup. Training cost ${\sim}1{,}105$ GPU-hours
(${\sim}53$--$74$\,kg\,CO$_2$e)~\cite{lacoste2019quantifying,dea_keyfigures2023}.

\paragraph{Metrics and statistical inference.}
\label{sec:metrics}
For class $c$ on exam $i$, $\mathrm{Dice} = 2\,\mathrm{TP} / (2\,\mathrm{TP} +
\mathrm{FP} + \mathrm{FN})$. Because its false-positive penalty is not scaled by
structure size -- the volume bias of ``\nameref{sec:confounders}'' -- we also report the
normalised Dice score (nDSC)~\cite{raina2023ndsc}, which rescales that term to the
dataset-mean volume (recovering Dice when a class is already average-sized), and HD95
(mm; lower better)~\cite{taha2015metrics}; all three are reported per class and
macro-averaged. For group disparities we use the binarized, rate-based metrics canonical
in this literature~\cite{parikh2025labelbias,bird2020fairlearn,eeoc1978uniform}: an exam
\emph{succeeds} when its score clears $\tau$ ($0.8$ for Dice/nDSC, $5$\,mm for HD95),
giving per-group success rates whose spread (DPD) and ratio (DIR, ${<}0.8$ flags adverse
impact) we read off a $\tau$-sweep. For significance we test the continuous scores --
Mann--Whitney~\cite{mann1947test} (sex, race) and Kruskal--Wallis~\cite{kruskal1952ranks}
(age) with rank-based effect sizes, FDR-corrected across the attribute$\times$metric
family~\cite{benjamini1995fdr} -- backed by bootstrap DIR intervals, a joint OLS on
$\mathcal{A}$, and permutation checks. These continuous tests are our \emph{primary}
instrument: at macro Dice ${\approx}0.89$ success rates sit near the ceiling and DIR is
near-degenerate, so the rank tests carry more power.

\paragraph{Experimental design (E1--E2).}
\label{sec:design}
Two experiments isolate where label provenance acts; each applies the full battery to
every attribute in $\mathcal{A}$. \textbf{E1} asks whether the deployed model is fair:
it audits $M_{\mathrm{mix}}$ on the full $228$-case test set against the dataset's own
(mixed) labels. \textbf{E2} asks whether the \emph{reference} distorts that verdict, on
the $76$ gold test cases scored against expert labels: it needs two references per image,
but CSpineSeg gives only one, so we generate the second as $M_{\mathrm{gold}}$'s
predictions (it never saw these $76$), mirroring how Zhou et al.\ produced the dataset's
silver labels~\cite{zhou2025cspineseg}, and score $M_{\mathrm{mix}}$'s \emph{same}
predictions against both the gold and this generated-silver ruler to isolate the pure
ruler effect.

\section{Experiments and Results}

For every regime nnU-Net selected the softmax-averaged ensemble of the 2D and 3D
full-resolution configurations, giving ten model-folds per regime (twenty total).
We report E1 and E2 (``\nameref{sec:design}'').

\subsection{Global Fairness Audit (E1)}
\label{sec:global}

On the full $228$-case test set, $M_{\mathrm{mix}}$ reaches a macro Dice of $0.947$
(vertebral body $0.959$, disc $0.935$), a median macro HD95 of $0.34$\,mm, and a
macro nDSC of $0.949$. On the $76$ gold-labelled cases -- the only ones
scored against expert references -- macro Dice is $0.897$, comparable to the $0.916$
Zhou et al.~\cite{zhou2025cspineseg} report on a different split with a smaller
training pool.

The model is fair by every measure: across sex, race, and age on all three metrics
every DIR exceeds $0.96$ and every BCa lower confidence bound clears the four-fifths
threshold of $0.80$.
None of the $63$ tests reach significance after FDR
correction (smallest $p_{\mathrm{fdr}} = 0.72$), effect sizes are negligible
throughout ($|r_{rb}| \le 0.14$, $\varepsilon^2 < 0.01$), and a joint OLS of macro
Dice on sex, race, and age explains $1.3\%$ of variance ($R^2 = 0.013$, $p = 0.43$).

\subsection{The Biased Ruler: Performance Inflation and False Confidence (E2)}
\label{sec:biased_ruler}

We now ask whether the choice of reference label can change a verdict. We take the
same model, the same $76$ gold-test images, and score them twice: against the
expert (gold) labels, and against $M_{\mathrm{gold}}$'s predictions on those images
-- the generated-silver ruler of ``\nameref{sec:design}''. The predictions are
identical, so any change in the verdict is a pure ruler effect.

\paragraph{Performance overestimation from label leakage.}
Against gold, $M_{\mathrm{mix}}$ scores $0.897$ macro Dice; against silver, the same
predictions score $0.973$ -- an inflation of nearly $8$ points. The silver ruler is
$M_{\mathrm{gold}}$'s output, and $M_{\mathrm{gold}}$ and $M_{\mathrm{mix}}$ were both
trained, in part, on the \emph{same} expert labels; two models fitted to overlapping
data come to resemble each other more than either resembles the ground truth, so they
share systematic errors and the silver ruler flatters them. This leakage is the
dataset's, not an artefact of our setup: CSpineSeg's published silver labels were
likewise generated by a model trained on its gold labels, so \emph{any} user who
validates against them overestimates performance, not just fairness auditors.

\paragraph{The verdict flip.}
Under the gold ruler, $0/63$ tests are FDR-significant (smallest
$p_{\mathrm{fdr}} = 0.13$); under the silver ruler, $11/63$ are, and every one is an
age comparison (\autoref{tab:biased_ruler}). The significance verdict reverses purely
from the choice of reference label.

\begin{figure}[t]
\centering
\includegraphics[width=0.9\textwidth]{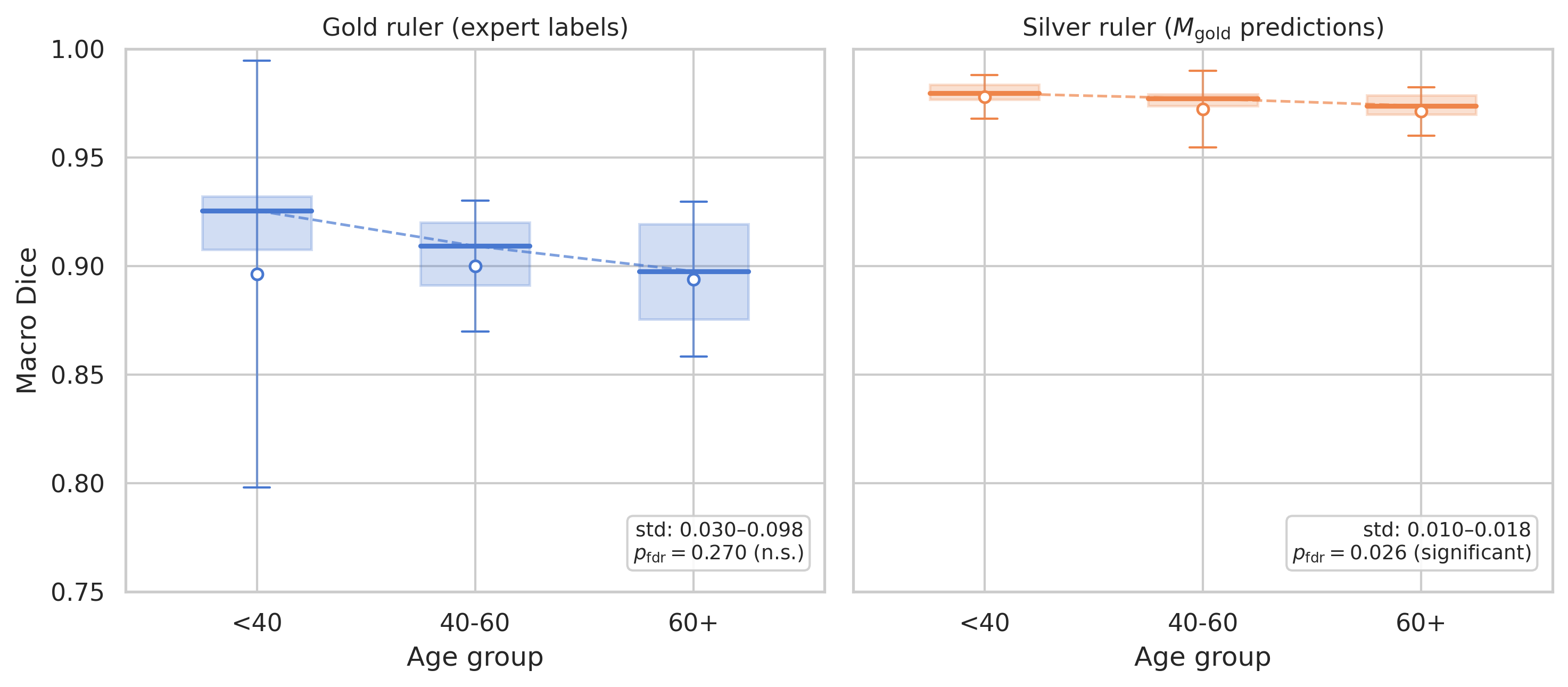}
\caption{The variance collapse behind the flip. Macro Dice by age bin for the same
$76$ predictions, scored against the gold ruler (left) and silver ruler (right). The
gradient runs the same direction on both (60+ worst), but gold's per-group SDs
($0.03$--$0.10$) bury it below FDR correction while silver's collapsed variance
($0.01$--$0.02$) exposes it. The ruler changed the test's power, not the patient.}
\label{fig:age_trend}
\end{figure}

\paragraph{Mechanism: variance collapse, not magnitude.}
The age trend is \emph{smaller} against silver: the worst-vs-best median spread
falls from $2.7$ Dice points (gold) to $0.6$ (silver). What changes is noise -- the
within-group standard deviation drops from $0.058$ to $0.014$ (${\sim}4\times$) -- so
the Kruskal--Wallis $H$ \emph{rises} from $7.0$ ($p_{\mathrm{fdr}} = 0.27$) to
$11.4$ ($p_{\mathrm{fdr}} = 0.026$). The rate-based DIR saturates at $1.0$ on silver
(zero failures above $\tau$) even as the continuous test turns significant; a
threshold sweep confirms the saturation. \autoref{fig:age_trend} shows the same
downward staircase on both rulers -- gold's wide boxes bury the signal, silver's
razor-thin boxes expose it. We interpret the two modes in
``\nameref{sec:discussion}''.

\begin{table}[t]
\centering
\caption{E2 biased ruler on the $76$ gold-test images (macro Dice). The same
$M_{\mathrm{mix}}$ predictions are scored against expert labels (gold) and
$M_{\mathrm{gold}}$'s predictions (silver). DIR/DPD are rate-based ($\tau = 0.8$);
$p_{\mathrm{fdr}}$ is Kruskal--Wallis after FDR correction. The silver ruler
saturates (DIR~$\equiv 1.0$) yet manufactures significance on age.}
\label{tab:biased_ruler}
\small
\begin{tabular*}{\textwidth}{@{\extracolsep{\fill}} l c c c c c c}
\hline\noalign{\smallskip}
 & \multicolumn{3}{c}{Gold ruler} & \multicolumn{3}{c}{Silver ruler} \\
\noalign{\smallskip}\cline{2-4}\cline{5-7}\noalign{\smallskip}
Grouping & DIR & DPD & $p_{\mathrm{fdr}}$ & DIR & DPD & $p_{\mathrm{fdr}}$ \\
\noalign{\smallskip}\hline\noalign{\smallskip}
Sex          & $1.000$ & $0.000$ & $0.997$ & $1.000$ & $0.000$ & $0.718$ \\
Race (W/B)   & $0.956$ & $0.044$ & $0.997$ & $1.000$ & $0.000$ & $0.328$ \\
Age (3 bins) & $0.952$ & $0.048$ & $0.270$ & $1.000$ & $0.000$ & $0.026^{\ast}$ \\
\noalign{\smallskip}\hline
\end{tabular*}

\smallskip
{\footnotesize $^{\ast}$\,FDR-significant at $\alpha = 0.05$. All $11$ significant
tests on the silver ruler are age groupings (three-bin and median split, Dice/nDSC).}
\end{table}

\section{Discussion}
\label{sec:discussion}

\paragraph{Two modes of ruler bias.}
Our silver ruler is $M_{\mathrm{gold}}$'s predictions, so scoring against it measures
inter-model \emph{agreement}, not accuracy. That agreement (${\sim}0.97$ Dice, with
the within-group SD collapsing ${\sim}4\times$ from $0.058$ to $0.014$) far exceeds
the ${\sim}0.89$ the same predictions reach against expert labels, and the cause is
label leakage: $M_{\mathrm{mix}}$ and the $M_{\mathrm{gold}}$ ruler are both trained,
in part, on the same expert labels, so two models fitted to overlapping data resemble
each other more than either resembles the truth. Because the Kruskal--Wallis
test's power scales with signal-to-noise, the same clinically negligible age gradient
that gold's noise buries clears FDR correction once the leakage-driven variance
collapse sharpens it: the ruler changed the test's power, not the patient. This
complements Parikh et al.~\cite{parikh2025labelbias}, whose \emph{independently
biased} silver ruler inflates the between-group gap (DIR $0.871 \to 0.815$) --
\emph{false magnitude}; our \emph{correlated} ruler instead shrinks the gap yet
manufactures significance by removing noise (\emph{false confidence}). Same
practical hazard -- \textbf{the verdict depends on the ruler}, opposite mechanism.

\paragraph{The age effect is intrinsic, not unfairness.}
The silver-ruler significance is an artefact, but the age \emph{trend} is real: it
runs the same direction on both rulers (60+ worst) and matches the degenerative
anatomy of ``\nameref{sec:confounders}'' -- intrinsic difficulty, not a label
artefact, and not unfairness either, the magnitudes being clinically negligible
(${\le}2.7$ Dice points, every group $>0.89$, all gold-ruler DIRs $>0.93$).

\paragraph{Limitations.}
(i)~a high-performance ceiling (all groups $>0.89$ Dice) leaves little room for
disparity to emerge; (ii)~a single institution, anatomy, sequence, and architecture,
with no external validation, and a small gold test set ($76$, ${\sim}20$ Black) that
limits power for race; (iii)~a correlated silver ruler -- a near-clone of the
model under test -- so we can demonstrate only false confidence, not false magnitude;
and (iv)~our silver ruler is a \emph{reconstruction} of the dataset's label-generation
process ($M_{\mathrm{gold}}$'s predictions, the original generator being unavailable), and
the gold and silver pools -- split by enrollment order, not at random -- differ modestly
in race and age, so any gold-vs-silver \emph{training} comparison confounds provenance
with composition (the E2 ruler result, on identical images, is immune).

\paragraph{Recommendations.}
(i)~\textbf{Report against expert labels.} On a mixed-provenance dataset, machine
labels leak information from the expert labels they were derived from, so validating
on them overestimates performance (${\sim}8$ Dice points here); the only safe
reference is the expert subset, which presupposes users are told which labels are
which. (ii)~\textbf{State where reference labels come from}: swapping expert for machine
labels moved our verdict from $0/63$ to $11/63$ significant on identical predictions.
(iii)~\textbf{Check ruler--model similarity}: scores much higher \emph{and} much tighter
against the ruler than against experts (${\sim}0.97$ vs.\ ${\sim}0.89$, $4\times$
variance collapse) signal a ruler too close to be an independent benchmark.
(iv)~\textbf{Report magnitude, not just significance}: the flagged age gap spans $0.6$
Dice points where every group exceeds $0.97$. Eliciting false magnitude with an
independent ruler, and a formal variance-collapse diagnostic, remain for future work.

\section{Conclusion}

We presented the first demographic fairness audit of cervical-spine MRI segmentation:
across sex, race, and age the deployed model is fair by every standard metric. The
decisive finding, however, is methodological and reaches beyond this anatomy. Because a
mixed-provenance dataset's automated silver labels come from a model trained on its expert-gold labels,
any new model trained on the same gold labels resembles the silver labels more than
expert truth -- so scoring against silver overestimates performance
(${\sim}8$ Dice points here) and can flip a fairness verdict by collapsing within-group
variance (\emph{false confidence}, the complement to Parikh et al.'s~\cite{parikh2025labelbias} \emph{false
magnitude}). 

We therefore \textbf{urge the community} to treat reference-label provenance as a
first-order, reportable variable: datasets should mark which labels are expert versus
machine-generated, and every fairness claim should name its reference's provenance, be
measured against expert labels where possible, and report a disparity's \emph{magnitude},
not only its significance. As a \textbf{limitation}, our evidence is from a single institution,
anatomy, and architecture in a high-performance regime; whether the same leakage distorts
harder tasks remains open.

%
%
\newpage
\bibliographystyle{splncs04}
\bibliography{mybibliography}
%
%
\end{document}